\documentclass[aps,prl,twocolumn,showpacs,preprintnumbers,amsmath,amssymb,refmerge,superscriptaddress]{revtex4}
\usepackage{longtable}
\usepackage{amsmath}
\usepackage{graphicx} % Include figure files
\usepackage{dcolumn}  % Align table columns on decimal point
\usepackage{color}

\renewcommand{\arraystretch}{1.1}

\newcommand{\mev}{\mathrm{MeV}}

\newcommand{\mevm}{\mathrm{MeV}/c^2}
\newcommand{\gev}{\mathrm{GeV}}

\newcommand{\gevm}{\mathrm{GeV}/c^2}
\newcommand{\gevms}{\mathrm{GeV}^2/c^4}
\newcommand{\ee}{e^{+} e^{-}}

\newcommand{\pipi}{\pi^{+}\pi^{-}}

\newcommand{\leplep}{\ell^{+}\ell^{-}}
\newcommand{\jp}{J/\psi}
\newcommand{\ch}{\chi_{c1}}
\newcommand{\chic}{\chi_{c1}}
\newcommand{\psp}{\psi'}
\newcommand{\psip}{\psi'}
\newcommand{\z}{Z(4430)^+}
\newcommand{\kst}{K^{*}}

\newcommand{\ks}{K^0_S}
\newcommand{\kstr}{K^{*}}
\newcommand{\B}{\bar{B}^{0}}

\newcommand{\Bp}{B^+}

\newcommand{\km}{K^-}
\newcommand{\pip}{\pi^{+}}

\newcommand{\pim}{\pi^{-}}

\newcommand{\Mbc}{M_{\rm bc}}
\newcommand{\de}{\Delta E}

\newcommand{\fb}{\mathrm{fb}^{-1}}
\newcommand{\br}{\mathcal{B}}
\newcommand{\lik}{\mathcal{L}}
\newcommand{\etal}{\em et al.}
\newcommand{\rt}{\rightarrow}

\newcommand{\mz}{(4443^{+15}_{-12}{^{+19}_{-13}})\,\mevm}
\newcommand{\gz}{(107^{+86}_{-43}{^{+74}_{-56}})\,\mev}
\newcommand{\bz}{(3.2^{+1.8}_{-0.9}{^{+5.3}_{-1.6}})\times10^{-5}}
\newcommand{\bkmpippsp}{(5.68\pm0.13\pm0.42)\times10^{-4}}
\newcommand{\bkstpsp}{(5.52^{+0.35}_{-0.32}{^{+0.53}_{-0.58}})\times10^{-4}}
\newcommand{\fl}{(44.8^{+4.0}_{-2.7}{^{+4.0}_{-5.3}})\%}

\begin{document}

\title{ \quad\\[0.5cm]
Dalitz analysis of $B\rt K\pip\psp$ decays and the $\z$}

\begin{abstract}
From a Dalitz plot analysis of $B\rt K\pip\psp$ decays, we find a
signal for $\z\rt\pip\psp$ with a mass $M=\mz$, width $\Gamma=\gz$,
product branching fraction
$\br(\B\rt\km\z)\times\br(\z\rt\pip\psp)=\bz$, and significance of
$6.4\,\sigma$ that agrees with previous Belle measurements based on
the same data sample.
In addition, we determine the branching fraction
$\br(B^0\rt\kst(892)^0\psp)=\bkstpsp$ and the fraction of
$\kst(892)^0$ mesons that are longitudinally polarized $f_L=\fl$.
These results are obtained from a $605\,\fb$ data sample that contains
657 million $B\bar{B}$ pairs collected near the $\Upsilon(4S)$
resonance with the Belle detector at the KEKB asymmetric energy
$e^+e^-$ collider.
\end{abstract}


%%% Paper:    Z(4430) in B -> K pi psi'
%%% Journal:  Physical Review D (Rapid Communication)
%%% Contacts: R. Mizuk (mizuk@itep.ru)
%%%           R. Chistov (chistov@itep.ru)
%%% Non-responding authors or those who said NO are commented out.
%%% ====================================================================
%%% Click the RELOAD button on your web browser to see the updated file.
%%% ====================================================================
%%% Use \input{author} to insert this material into your latex file.
%%%%% Force institutions to appear in alphabetical order when typeset.
\affiliation{Budker Institute of Nuclear Physics, Novosibirsk}
\affiliation{Chiba University, Chiba}
\affiliation{University of Cincinnati, Cincinnati, Ohio 45221}
\affiliation{T. Ko\'{s}ciuszko Cracow University of Technology, Krakow}
\affiliation{Department of Physics, Fu Jen Catholic University, Taipei}
%%%\affiliation{Justus-Liebig-Universit\"at Gie\ss{}en, Gie\ss{}en}
\affiliation{The Graduate University for Advanced Studies, Hayama}
\affiliation{Gyeongsang National University, Chinju}
\affiliation{Hanyang University, Seoul}
\affiliation{University of Hawaii, Honolulu, Hawaii 96822}
\affiliation{High Energy Accelerator Research Organization (KEK), Tsukuba}
%%%\affiliation{Hiroshima Institute of Technology, Hiroshima}
%%%\affiliation{University of Illinois at Urbana-Champaign, Urbana, Illinois 61801}
\affiliation{Institute of High Energy Physics, Chinese Academy of Sciences, Beijing}
\affiliation{Institute of High Energy Physics, Vienna}
\affiliation{Institute of High Energy Physics, Protvino}
%%%\affiliation{Institute of Mathematical Sciences, Chennai}
%%%\affiliation{INFN - Sezione di Torino, Torino}
\affiliation{Institute for Theoretical and Experimental Physics, Moscow}
\affiliation{J. Stefan Institute, Ljubljana}
\affiliation{Kanagawa University, Yokohama}
%%%\affiliation{Institut f\"ur Experimentelle Kernphysik, Universit\"at Karlsruhe, Karlsruhe}
\affiliation{Korea University, Seoul}
%%%\affiliation{Kyoto University, Kyoto}
\affiliation{Kyungpook National University, Taegu}
\affiliation{\'Ecole Polytechnique F\'ed\'erale de Lausanne (EPFL), Lausanne}
\affiliation{Faculty of Mathematics and Physics, University of Ljubljana, Ljubljana}
\affiliation{University of Maribor, Maribor}
\affiliation{University of Melbourne, School of Physics, Victoria 3010}
\affiliation{Nagoya University, Nagoya}
\affiliation{Nara Women's University, Nara}
\affiliation{National Central University, Chung-li}
\affiliation{National United University, Miao Li}
\affiliation{Department of Physics, National Taiwan University, Taipei}
\affiliation{H. Niewodniczanski Institute of Nuclear Physics, Krakow}
\affiliation{Nippon Dental University, Niigata}
\affiliation{Niigata University, Niigata}
\affiliation{University of Nova Gorica, Nova Gorica}
\affiliation{Novosibirsk State University, Novosibirsk}
\affiliation{Osaka City University, Osaka}
%%%\affiliation{Osaka University, Osaka}
\affiliation{Panjab University, Chandigarh}
%%%\affiliation{Peking University, Beijing}
\affiliation{Princeton University, Princeton, New Jersey 08544}
%%%\affiliation{RIKEN BNL Research Center, Upton, New York 11973}
\affiliation{Saga University, Saga}
\affiliation{University of Science and Technology of China, Hefei}
\affiliation{Seoul National University, Seoul}
\affiliation{Shinshu University, Nagano}
\affiliation{Sungkyunkwan University, Suwon}
\affiliation{University of Sydney, Sydney, New South Wales}
\affiliation{Tata Institute of Fundamental Research, Mumbai}
%%%\affiliation{Toho University, Funabashi}
\affiliation{Tohoku Gakuin University, Tagajo}
%%%\affiliation{Tohoku University, Sendai}
\affiliation{Department of Physics, University of Tokyo, Tokyo}
\affiliation{Tokyo Institute of Technology, Tokyo}
\affiliation{Tokyo Metropolitan University, Tokyo}
\affiliation{Tokyo University of Agriculture and Technology, Tokyo}
%%%\affiliation{Toyama National College of Maritime Technology, Toyama}
\affiliation{IPNAS, Virginia Polytechnic Institute and State University, Blacksburg, Virginia 24061}
\affiliation{Yonsei University, Seoul}
  \author{R.~Mizuk}\affiliation{Institute for Theoretical and Experimental Physics, Moscow} % ITEP
  \author{I.~Adachi}\affiliation{High Energy Accelerator Research Organization (KEK), Tsukuba} % KEK
  \author{H.~Aihara}\affiliation{Department of Physics, University of Tokyo, Tokyo} % Tokyo
  \author{K.~Arinstein}\affiliation{Budker Institute of Nuclear Physics, Novosibirsk}\affiliation{Novosibirsk State University, Novosibirsk} % BINP
% \author{T.~Aso}\affiliation{Toyama National College of Maritime Technology, Toyama} % Toyama
% \author{V.~Aulchenko}\affiliation{Budker Institute of Nuclear Physics, Novosibirsk}\affiliation{Novosibirsk State University, Novosibirsk} % BINP
  \author{T.~Aushev}\affiliation{\'Ecole Polytechnique F\'ed\'erale de Lausanne (EPFL), Lausanne}\affiliation{Institute for Theoretical and Experimental Physics, Moscow} % ITEP
% \author{T.~Aziz}\affiliation{Tata Institute of Fundamental Research, Mumbai} % Tata
% \author{S.~Bahinipati}\affiliation{University of Cincinnati, Cincinnati, Ohio 45221} % Cincinnati
  \author{A.~M.~Bakich}\affiliation{University of Sydney, Sydney, New South Wales} % Sydney
  \author{V.~Balagura}\affiliation{Institute for Theoretical and Experimental Physics, Moscow} % ITEP
% \author{Y.~Ban}\affiliation{Peking University, Beijing} % Peking
  \author{E.~Barberio}\affiliation{University of Melbourne, School of Physics, Victoria 3010} % Melbourne
  \author{A.~Bay}\affiliation{\'Ecole Polytechnique F\'ed\'erale de Lausanne (EPFL), Lausanne} % Lausanne
% \author{I.~Bedny}\affiliation{Budker Institute of Nuclear Physics, Novosibirsk}\affiliation{Novosibirsk State University, Novosibirsk} % BINP
  \author{K.~Belous}\affiliation{Institute of High Energy Physics, Protvino} % Protvino
% ...
% BINP
\collaboration{The Belle Collaboration}

\maketitle

\tighten

{\renewcommand{\thefootnote}{\fnsymbol{footnote}}}
\setcounter{footnote}{0}

\section{Introduction}

In a paper on the B meson decay process $B\rt
K\pip\psp$~\cite{conjugate}, the Belle Collaboration~\cite{Z4430}
reported the observation of a distinct and relatively narrow peak in
the $\pip\psp$ mass spectrum near $M(\pip\psp)\simeq4430\,\mevm$.
The analysis was performed by excluding the events in the $M(K\pip)$
regions of the $\kst(892)$ and $\kst(1430)$, and fitting the
one-dimensional $M(\pip\psp)$ distribution. The fit gave a resonance
mass and width of
$M=(4433\pm4\pm2)\,\mevm$ and
$\Gamma=(45^{+18}_{-13}{^{+30}_{-13}})\,\mev$, where the first
uncertainty is statistical and the second is systematic; the
significance of the resonance was $6.5\sigma$.
If this peak, called the $\z$, is interpreted as a meson state, then
it must have an exotic structure; its minimal quark content is
$|c\bar{c}u\bar{d}\rangle$.

The $\z$ observation motivated a subsequent Belle study of the process
$\B\rt\km\pip\chic$, where a doubly peaked structure was observed in
the $\pip\chic$ invariant mass distribution~\cite{Z4040}. In this
channel, the observed structure is rather wide, therefore a full
Dalitz plot analysis was used in order to establish that the observed
peaks could be unambiguously associated with dynamics in the
$\pip\chic$ channel. If these peaks, called the $Z(4040)$ and
$Z(4240)$, are attributed to meson states, a minimal four-quark
substructure is required.

A recently reported study of $B\rt K\pip\psip$ decays by the BaBar
Collaboration~\cite{BaBar_Z4430} did not find a significant signal for
$\z\rt\pip\psip$; the reported significance is at the $1.9\,\sigma -
3.1\,\sigma$ level. The BaBar sample of $B\rt K\pip\psip$ decays is
about 85\% the size of the corresponding Belle data sample.

In this paper we present a reanalysis of the Belle $B\rt K\pip\psp$
data sample using a Dalitz plot formalism. 

The Belle detector~\cite{BELLE_DETECTOR} is a large-solid-angle
magnetic spectrometer that operates at the KEKB asymmetric-energy
$\ee$ collider~\cite{kekb}. A data sample corresponding to an
integrated luminosity of $605\,{\rm fb}^{-1}$ collected at the
$\Upsilon(4S)$ resonance and containing 657 million $B\bar{B}$ pairs
is used.
A GEANT-based Monte Carlo (MC) simulation~\cite{geant} is used to
model the response of the detector.

\section{Event Selection}

We select events of the type $\B\rt\km\pip\psp$ and
$\Bp\rt\ks\pip\psp$, where the $\psp$ decays either to $\leplep$ or
$\pipi\jp$ with $\jp\rt\leplep$ ($\ell = e$ or $\mu$),
$\ks\rt\pip\pim$. We use the same selection criteria as in
Ref.~\cite{Z4430}.
In particular, we identify $B$ mesons using the beam-energy
constrained mass $\Mbc=\sqrt{E_{\rm beam}^2-p_B^2}$ and the energy
difference $\de=E_{\rm beam}-E_B$, where $E_{\rm beam}$ is the
center-of-mass (c.m.)\ beam energy, $p_B$ is the vector sum of the
c.m.\ momenta of the $B$ meson decay products and $E_B$ is their
c.m.\ energy sum. We select events with $|\Mbc-m_B|<7.1\,\mevm$ ($m_B$
is the world-average $B$-meson mass~\cite{PDG}) and $|\de|<34\,\mev$,
which are both $\pm2.5\,\sigma$ windows around the peak values. To
model combinatorial backgrounds, we use events that are in the $\Mbc$
signal region and the $\de$ sidebands defined as
$|\de\pm70\,\mev|<34\,\mev$.
To improve the definition of the Dalitz plot boundaries for both
signal and sideband events, we perform a mass-constrained fit to the
$B$ candidates from both regions. 
Simulations of the two $\psp$ decay modes indicate that the
experimental resolution for $M(\pip\psp)$ is $\sigma=2.5\,\mevm$ for
both modes.

\section{Dalitz plot distribution}

\begin{figure}[htbp]
\includegraphics[width=8cm]{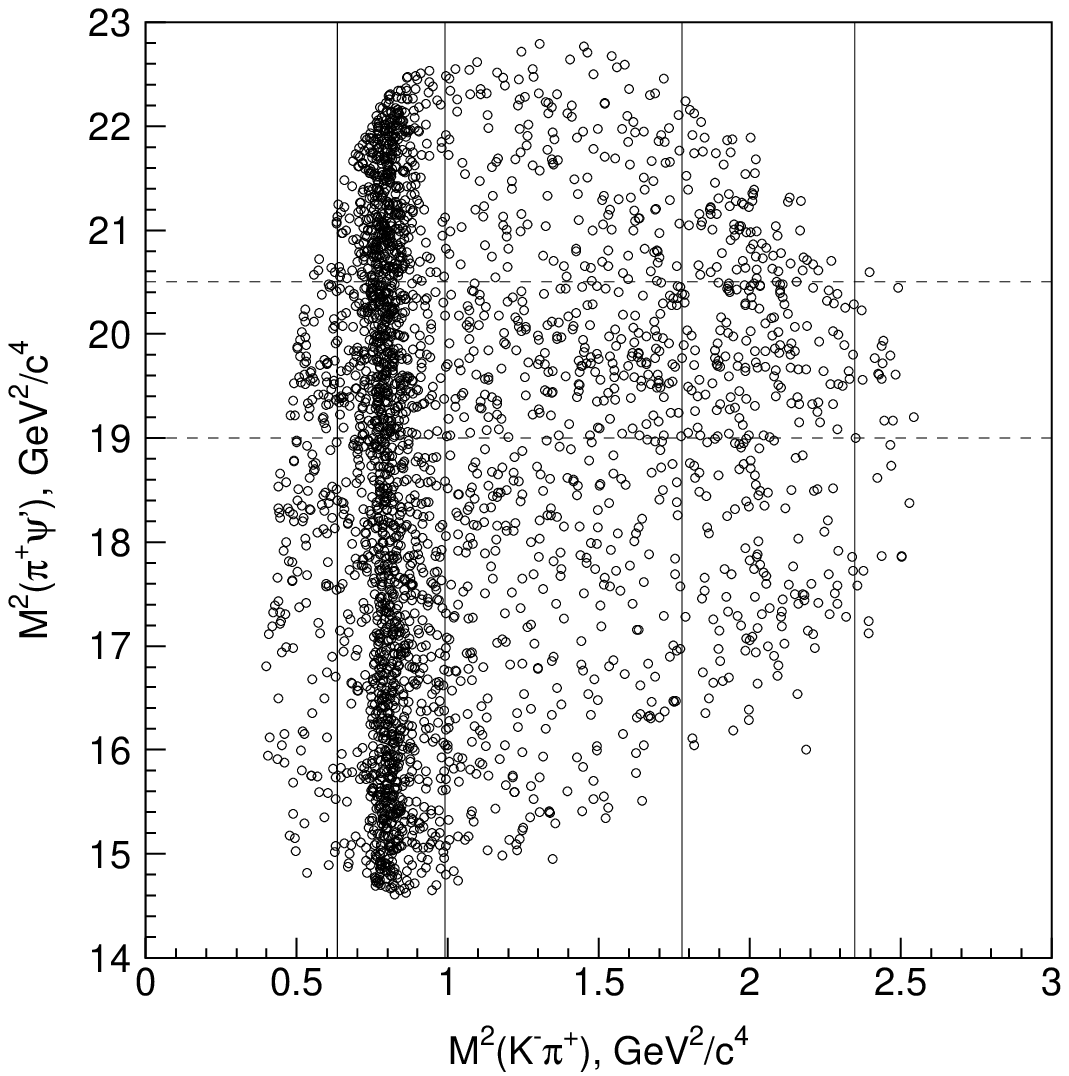}
\caption{ The $B\rt K\pip\psp$ Dalitz plot for the $\de$ signal
  region. The solid (dashed) lines delimit the five vertical (three
  horizontal) slices that are used to present the fit results in
  subsequent figures. The coordinates of the vertical lines are
  $M^2(K\pip)=(0.796)^2\,\gevms$, $(0.996)^2\,\gevms$,
  $(1.332)^2\,\gevms$ and $(1.532)^2\,\gevms$; the coordinates of the
  horizontal lines are $M^2(\pip\psp)=19.0\,\gevms$ and
  $20.5\,\gevms$. }
\label{fig:dp}
\end{figure}

We sum the Dalitz distributions for $\B\rt\km\pip\psp$ and
$\Bp\rt\ks\pip\psp$ candidates. Due to the mass difference between
$\km$ and $\ks$ the corresponding Dalitz plots have slightly different
boundaries. We find that this has a negligible effect on the results
of the Dalitz analysis. The Dalitz plot for the $\de$ signal region is
shown in Fig.~\ref{fig:dp}. Here vertical bands corresponding to the
$\kst(892)$ and the $\kst(1430)$ are evident. The horizontal cluster
of entries in the vicinity of $M^2(\pip\psp)\sim20\,\gevms$
constitutes the $\z$ signal reported in Ref.~\cite{Z4430}.

In the following, we illustrate the results of different fits using
projected histograms of the slices of the Dalitz plot indicated by the
vertical solid lines and horizontal dashed lines shown in
Fig.~\ref{fig:dp}. The three horizontal slices correspond to
$M(\pip\psp)$ regions below, around and above the $\z$ mass region.
The five vertical slices distinguish the $\kst(892)$ and
$M(K\pip)\simeq1.4\gevm$ regions and bands above, below and in between
them. The sum of the latter three projections corresponds to the
$\kst$ veto used in Ref.~\cite{Z4430}.

\section{Formalism of the Dalitz analysis}

The decay $B\rt K\pip\psp$ with the $\psp$ reconstructed in the
$\leplep$ decay mode is described by four variables (assuming the
width of the $\psp$ to be negligible). These are taken to be
$M(\pip\psp)$, $M(K\pip)$, the $\psp$ helicity angle ($\theta$) and
the angle between the $\psp$ production and decay planes ($\phi$). In
this analysis we integrate over the angular variables $\theta$ and
$\phi$. The MC indicates that the reconstruction efficiency is almost
uniform over the full $\phi$ angular range; after integration over
this angle the contribution from interference between the different
$\psp$ helicity states is negligibly small. This allows the $\psp$ to
be treated as a stable particle in the Dalitz analysis.

In the $\psp\rt\pipi\jp$ channel, the $\psp$ is likewise treated as
stable. The $\pipi$ system in this decay is predominantly in an
$S$-wave~\cite{BES_pipijp}; in this limit, the $\psp$ and $\jp$
helicity states are the same, and we again find negligible
interference contributions after integration over decay angles.
Thus, our approach is the same as in the Dalitz analysis of the
$\B\rt\km\pip\ch$ decays in Ref.~\cite{Z4040}.

The amplitude for the three-body decay $B\rt K\pip\psp$ is a sum over
different quasi-two-body modes; resonances are described by
relativistic Breit-Wigner functions with angular dependence.
As the default fit model, we include all known low-lying $K\pip$
resonances [the $\kappa$ or $\kst(800)$, and the $\kst(892)$,
  $\kst(1410)$, $\kst_0(1430)$, $\kst_2(1430)$, and $\kst(1680)$] and
a single exotic $\pip\psp$ resonance.
In addition to the physics model, the fit function includes a
background term derived from the $\de$ sidebands and is modulated by
the MC-determined experimental efficiency. 
The MC sample is generated using the world-average $\psp$ branching
fractions~\cite{PDG} while to fix the relative fractions of the $B^0$
and $B^+$ contributions we use isospin symmetry.
The Dalitz plots for the $\de$ sideband and the MC sample are
smoothed.
The expression for the amplitudes, signal component of the fit
function, and other details of the fitting procedure are the same as
used in the analysis described in Ref.~\cite{Z4040}.

\section{Fit results}

\begin{figure}[p!]
\includegraphics[width=8cm]{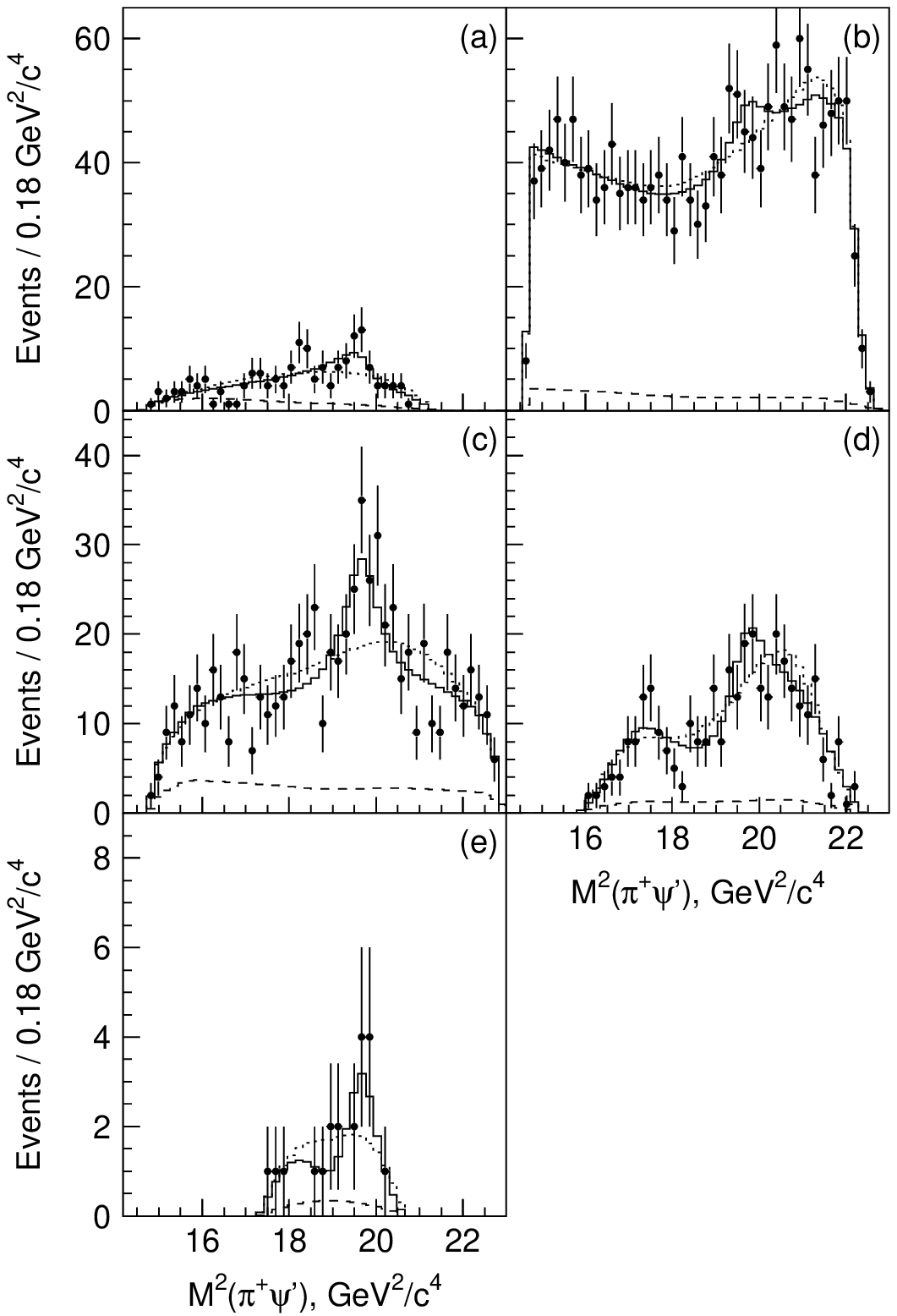}
\includegraphics[width=8cm]{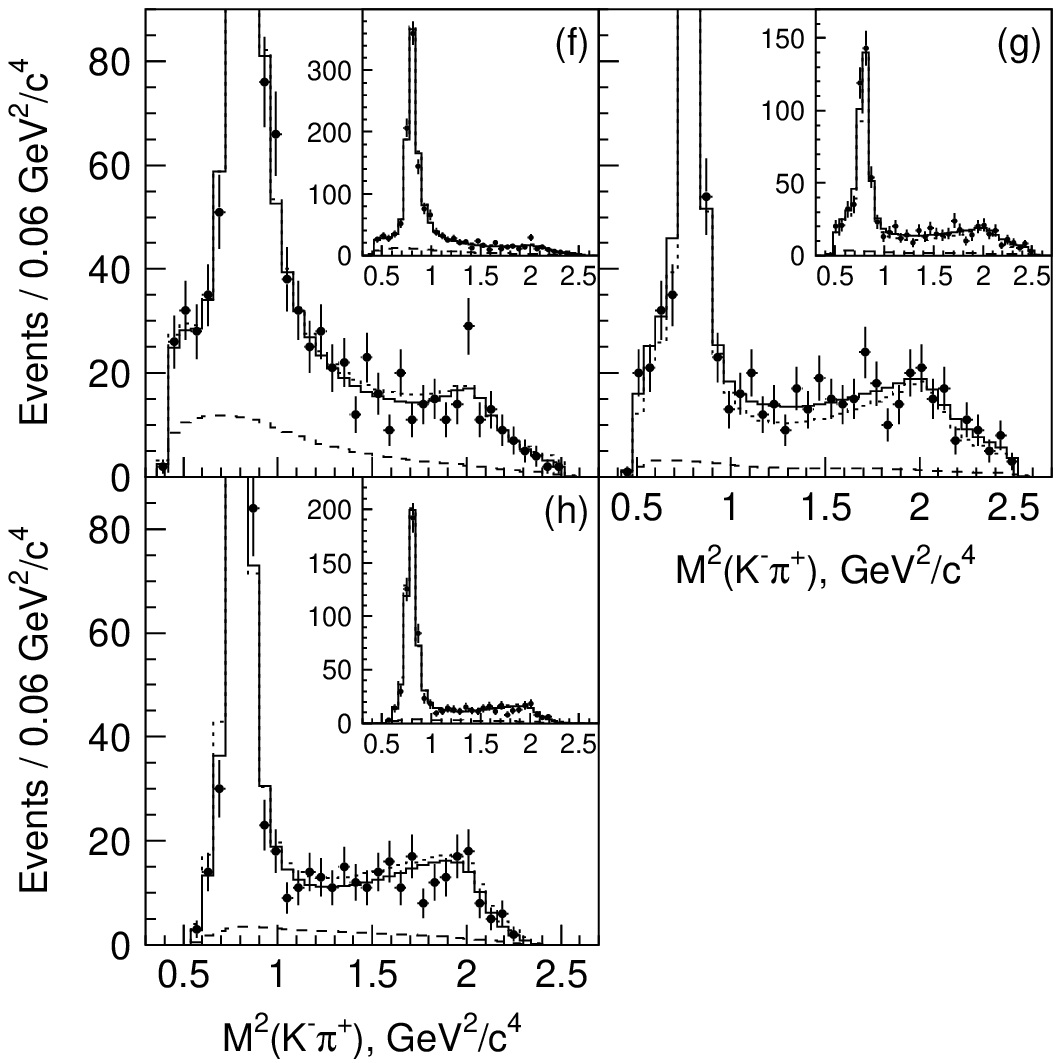}
\caption{ Dalitz plot projections for the slices defined in
  Fig.~\ref{fig:dp}: (a)-(e) correspond to vertical slices from left
  to right, (f)-(h) correspond to horizontal slices from bottom to
  top; in (f)-(h), plots including the full vertical scale are shown
  inset. The points with error bars represent data, the solid (dotted)
  histograms are the fit results for the default model that includes
  all low-lying $K\pi$ resonances and a single (without any) $\pip\psp$
  state, and the dashed histograms represent the background. }
\label{fig:py_px}
\end{figure}

\begin{figure}[htbp]
\includegraphics[width=6cm]{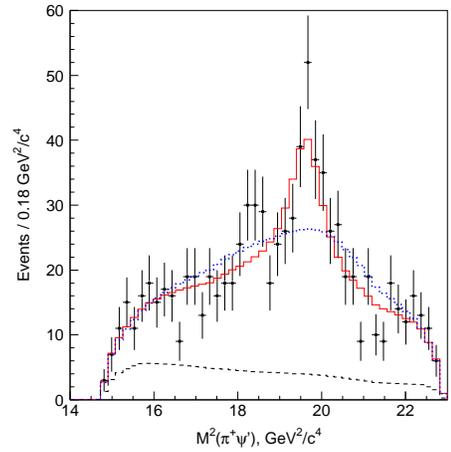}
\caption{ The Dalitz plot projection with the $\kst$ veto applied.
  The points with error bars represent data, the solid (dotted)
  histogram is the Dalitz plot fit result for the fit model with all
  $K\pi$ resonances and a single (without any) $\pip\psp$ state, and the
  dashed histogram represents the background. }
\label{fig:sum}
\end{figure}

The eight projected Dalitz plot slices with fit results for the
default model superimposed are shown in Fig.~\ref{fig:py_px}.
The $\z$ signal is most clearly seen in the third vertical slice.
The sum of the 1st, 3rd and 5th vertical slices ({\it i.e.} a Dalitz
plot projection with the $\kst$ veto applied) is shown in
Fig.~\ref{fig:sum}.
The $\pip\psp$ resonance parameters determined from the fit are
$M=(4443^{+15}_{-12})\,\mevm$ and $\Gamma=(107^{+86}_{-43})\,\mev$.
The central values agree well with the parameters reported in
Ref.~\cite{Z4430}, while the errors are somewhat larger. The
statistical significance, calculated from the change in $2\log \lik$
when the $\z$ is included in the fit (taking the added degrees of
freedom into account) is $6.4\,\sigma$. The fit fractions and
significances for all of the components are listed in
Table~\ref{ff-sig}.
\begin{table}[htb]
\caption{The fit fractions and significances of all contributions for
  the fit models with the default set of $K\pip$ resonances and a
  single $\pip\psp$ resonance.}
\label{ff-sig}
\renewcommand{\arraystretch}{1.5}
\begin{ruledtabular}
\begin{tabular}{c|cc}
Contribution & Fit fraction (\%) & Significance \\
\hline
$\z$           & $ 5.7^{+3.1}_{-1.6}$   & $6.4\,\sigma$ \\
$\kappa$       & $ 4.1^{+3.4}_{-1.1}$   & $1.5\,\sigma$ \\
$\kst(892)$     & $ 64.8^{+3.8}_{-3.5}$ & large \\
$\kst(1410)$    & $ 5.5^{+8.8}_{-1.5}$  & $0.5\,\sigma$ \\
$\kst_0(1430)$  & $ 5.3 \pm 2.6$        & $1.3\,\sigma$ \\
$\kst_2(1430)$  & $ 5.5^{+1.6}_{-1.4}$   & $3.1\,\sigma$ \\
$\kst(1680)$    & $ 2.8^{+5.8}_{-1.0}$  & $1.2\,\sigma$ \\
\end{tabular}
\end{ruledtabular}
\end{table}
The confidence level (C.L.) of the fit model with (without) the $\z$
is 36\% (0.1\%). The C.L.'s are determined using ensembles of the MC
simulated experiments.

To study the model dependence, we consider a variety of other fit
hypotheses. These include: successively removing each $\kst$ resonance
component; adding, for each case, a non-resonant phase-space term;
relaxing the constraints on the $\kappa$ mass and width; replacing the
$\kappa$ with the LASS group's parameterization for the $K\pi$
$S$-wave amplitude~\cite{LASS}, and including another $J=1$ ($J=2$)
$\kst$ resonance with mass and width left as free parameters.
The lowest $\z$ significance of $5.4\,\sigma$ corresponds to the model
with a non-resonant phase-space term and a new $J=2$ $\kst$ resonance.
We treat the maximum variation of the $\z$ parameters from these
different fit models as the systematic uncertainty. The resulting
uncertainty estimates are given in the first row of
Table~\ref{z_syst}.

We find the uncertainty due to the variation of the $r$ parameter in
the Blatt-Weisskopf form factors~\cite{blatt-weisskopf} to be
negligible. The contribution of the uncertainties in the mass and
width of intermediate $\kst$ resonances that are fixed in the fit is
also found to be negligible.

We vary the assumption about the value of the $B$ decay orbital
angular momentum ($L$) for those cases where several possibilities
exist. The resulting uncertainties are given in the second row in
Table~\ref{z_syst}.

In the fits described above, the spin of the $\z$ is assumed to be
zero. We find that the $J=1$ assumption does not significantly improve
the fit quality. The variations in the $\z$ parameters for the
different spin assignments are considered as systematic uncertainties
and are listed in the third row in Table~\ref{z_syst}.

We consider alternative smoothing procedures for $\de$ sidebands and
MC samples. The corresponding variation of the $\z$ parameters are
given in the fourth row in Table~\ref{z_syst}.

To obtain the total systematic uncertainties, the values given in
Table~\ref{z_syst} are added in quadrature. The resulting mass, width
and fit fraction are $M=\mz$, $\Gamma=\gz$,
$f=(5.7^{+3.1}_{-1.6}{^{+9.4}_{-2.7}})\%$. 

\begin{table}[htb]
\caption{Systematic uncertainties in the $\z$ mass, width and fit
  fraction due to various sources. }
\label{z_syst}
\renewcommand{\arraystretch}{1.5}
\begin{ruledtabular}
\begin{tabular}{l|ccc}
        & $M,\,\mevm$ & $\Gamma,\,\mev$ & Fit fraction, \% \\
\hline
Fit model           & $^{+14}_{-13}$ & $^{+56}_{-52}$ & $^{+3.6}_{-2.7}$ \\
$L$ assignment       & $^{+8}_{-0}$ & $^{+44}_{-0}$ & $^{+2.0}_{-0.0}$ \\
$Z$ spin assignment & $^{+9}_{-0}$  & $^{+8}_{-0}$ & $^{+8.4}_{-0}$ \\
Smoothing procedure  & $^{+4}_{-3}$  & $^{+17}_{-23}$ & $^{+0.5}_{-0.2}$ \\
\end{tabular}
\end{ruledtabular}
\end{table}

\section{Other fits}

\begin{figure}[htbp]
\includegraphics[width=8cm]{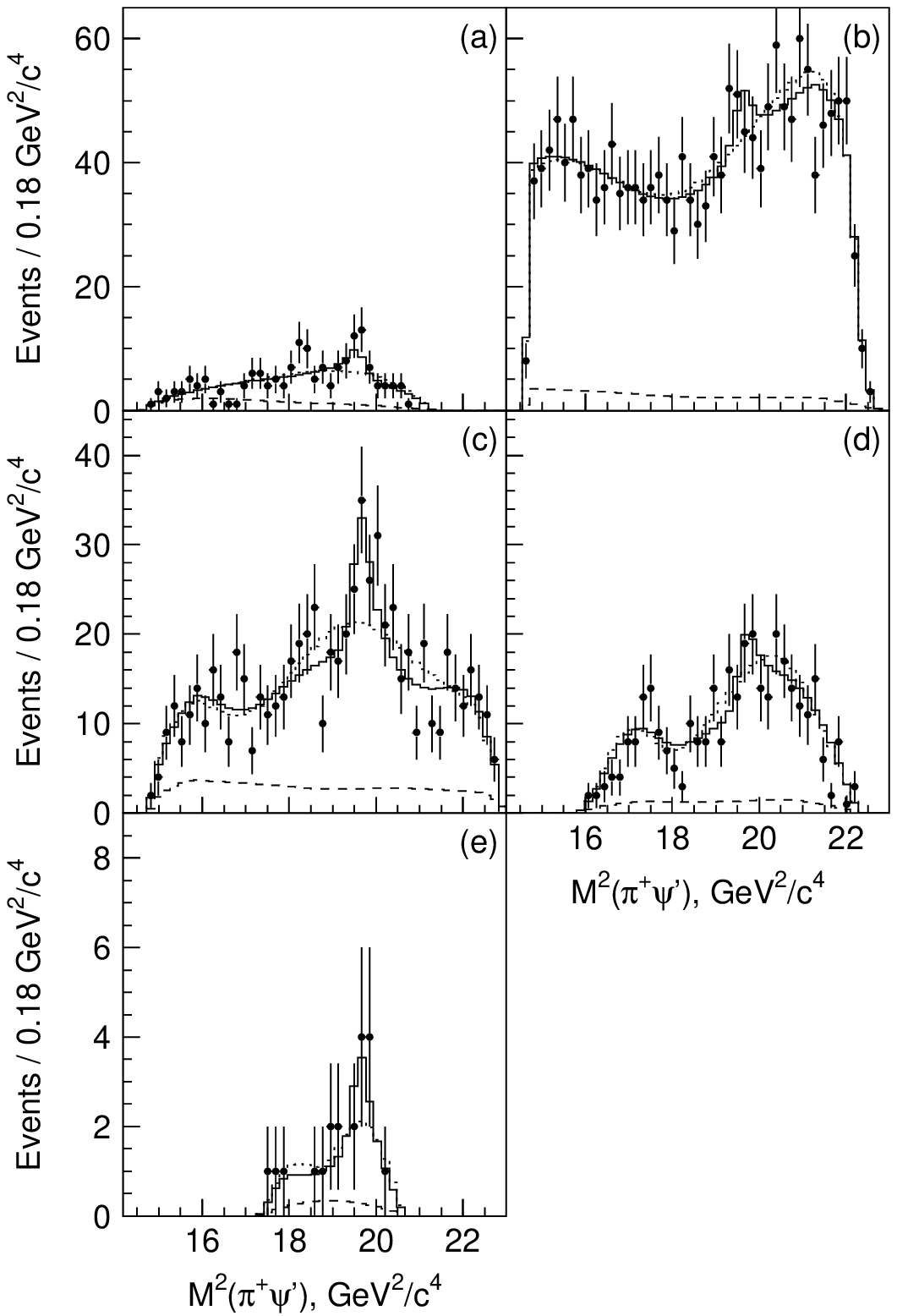}
\includegraphics[width=8cm]{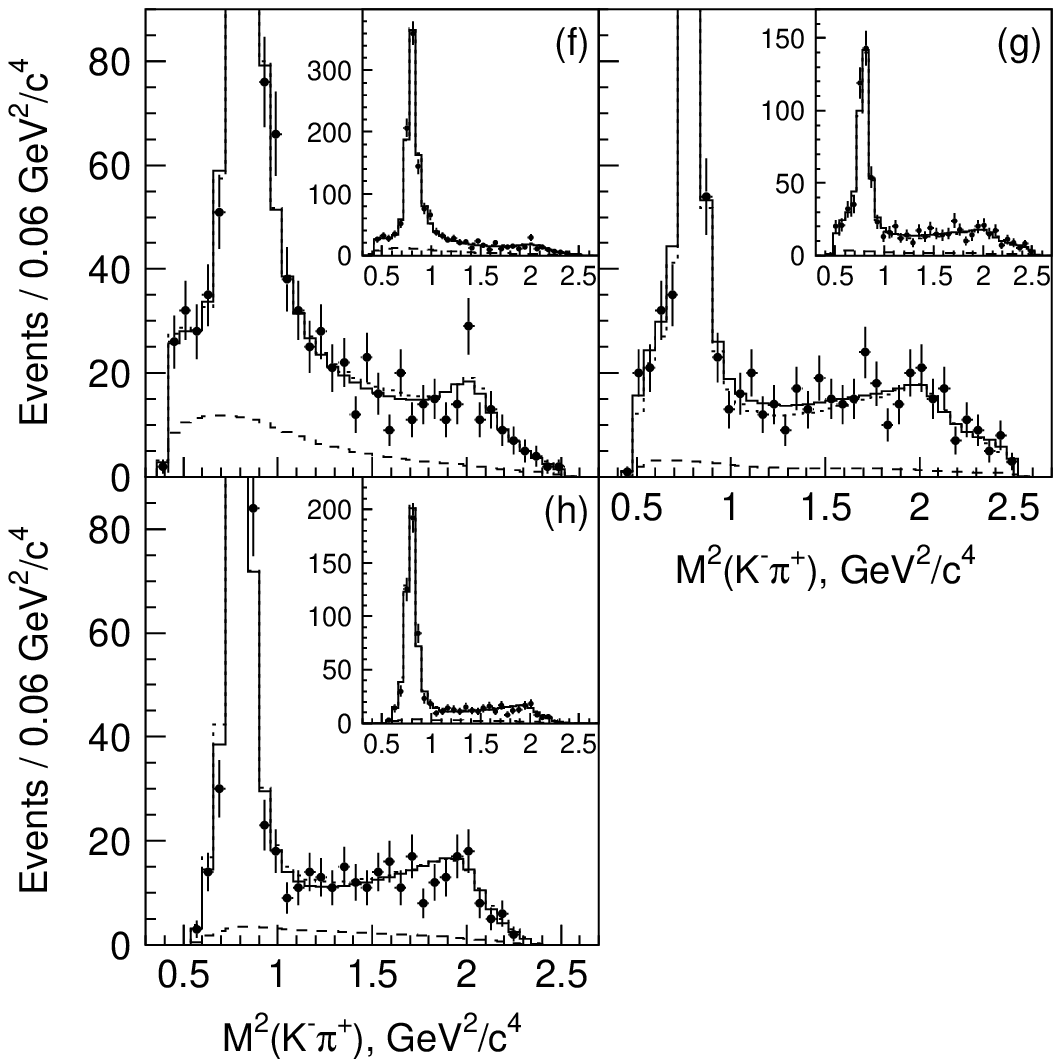}
\caption{ Dalitz plot projections as described for
  Fig.~\ref{fig:py_px}. The points with error bars represent data, the
  solid (dotted) histograms are the fit results for the model that
  includes all low-lying $K\pi$ resonances as well as the
  $\kst_3(1780)$ and a single (without any) $\pip\psp$ state, and the
  dashed histograms represent the background. }
\label{fig:add1780_py_px}
\end{figure}

In principle, more complex mass structures can be produced by
reflections from higher $K\pip$ partial waves. To examine this, we
perform the Dalitz plot fit with a $\kst_3(1780)$ resonance term added
to the default model (see Fig.~\ref{fig:add1780_py_px}).
In this case, the $\z$ signal persists with mass and width within
$1\,\sigma$ of their default model values and with a statistical
significance of $4.7\,\sigma$. However, the $\kst_3(1780)$ fit
fractions --- 6.8\% and 6.6\% for the $\z$ and non-$\z$ hypotheses,
respectively --- are very large for a resonance with a peak mass that
is $\sim 180\,\mevm$ ($\simeq 1.2\Gamma_{\kstr_3(1780)}$) above the
kinematic limit for $B\rt K\pip\psp$ decays and for which only a small
portion of the low-mass tail of the resonance is accessible. Moreover
$B\rt\kst_3(1780)\psp$ decay has an $L\ge2$ centrifugal barrier and
the $\kst_3(1780)\rt K\pi$ branching fraction is only
$(18.8\pm1.0)\%$~\cite{PDG}.
For these reasons, the $4.7\,\sigma$ significance estimate from this
fit model is likely to be an underestimate of the real value.
Studies of other $B$ decays where the $\kst_3(1780)$ can contribute
(e.g. $B\rt K\pip\jp$ and $B\rt\kst\pip\psp$) may provide further
insight.
The C.L. of the default model with an additional contribution from the
$\kst_3(1780)$ is 58\% (6\%) for the $\z$ (non-$\z$) hypothesis.
The significant $\z$ contribution is concentrated in a small area of
the Dalitz plot.

If a second $Z$ state is added to the fit, we find a mass
$M\sim4.3\,\gevm$ and a width $\Gamma\sim0.2\,\gev$, with a
significance of $3.9\,\sigma$.

Angular distributions for $\psp$ decays can be predicted based on the
Dalitz plot fit results and therefore provide a useful cross-check
(see Ref.~\cite{Z4040} for details). We find good agreement between
data and predictions for various fit models. The statistics are not
sufficient to discriminate between models in our approach.

\section{Branching fractions}

To measure branching fractions we use only $\B\rt\km\pip\psp$ decays.
The yields of these decays with the $\psp$ reconstructed in the
$\leplep$ and $\pipi\jp$ channels are found from fits to the $\de$
distributions to be $1089\pm34$ and $1166\pm37$, respectively.

To determine the experimental efficiency, we used the phase-space MC
events weighted according to the results of the Dalitz plot fit. The
efficiencies are $(19.2\pm1.4)\%$ and $(8.2\pm0.7)\%$ for
$\psp\rt\leplep$ and $\psp\rt\pipi\jp$ channels, respectively.  The
uncertainties include the dependence on the Dalitz plot model (0.1\%);
data and MC differences for track reconstruction ($1\%$ per track),
and particle identification (4\% for the $\km\pip$ pair and 4.2\% for
$\leplep$); and MC statistics (0.6\%). The uncertainties from
different sources are added in quadrature. The efficiencies are
corrected for the difference in lepton identification performance in
data compared to MC, ($-4.5\pm 4.2)\%$, determined from
$\jp\rt\leplep$ and $e^+e^-\rt e^+e^-\leplep$ control samples.

Using $(656.7\pm8.9)\times10^6/2$ as the number of $B^0\B$ pairs and
world-average values for the intermediate branching
fractions~\cite{PDG}, we determine $\br(\B\rt\km\pip\psp)=\bkmpippsp$.
This value is in agreement with BaBar's value
$(5.57\pm0.16)\times10^{-4}$~\cite{BaBar_Z4430} (statistical error
only).
The branching fractions calculated for the two $\psp$ decay channels
are in good agreement: $(5.73\pm0.18)\times10^{-4}$ and
$(5.62\pm0.18)\times10^{-4}$ (statistical errors only).
The systematic error includes contributions from the uncertainties in
the efficiencies and the branching fractions of the intermediate
resonances. 

Based on the $\z$ fit fraction we find a product branching fraction
$\br(\B\rt\km\z)\times\br(\z\rt\pip\psp)=\bz$. This is in agreement
with the previous Belle result~\cite{Z4430} and consistent with the
BaBar upper limit of $3.1\times10^{-5}$~\cite{BaBar_Z4430}.

The dominant feature of the $B\rt K\pip\psp$ decay process is the
$B\rt\kst(892)\psp$ intermediate state. Using the fit fraction from
Table~\ref{ff-sig}, and evaluating systematic errors using the same
procedure as for the $\z$ measurements, we determine the branching
fraction $\br(B^0\rt\kst(892)^0\psp)=\bkstpsp$ and the fraction of
$\kst(892)^0$ mesons that are longitudinally polarized $f_L=\fl$. The
branching fraction is somewhat below the world-average value of
$(7.2\pm0.8)\times10^{-4}$~\cite{PDG}; the longitudinal polarization
fraction agrees with the CLEOII result of $0.45\pm0.11\pm0.04$ and has
better precision~\cite{fl_cleo}.

\section{Conclusions}

From a Dalitz plot analysis of $B\rt K\pip\psp$ decays, we find a
signal for $\z\rt\pip\psp$ with a mass $M=\mz$, width $\Gamma=\gz$ and
product branching fraction
$\br(\B\rt\km\z)\times\br(\z\rt\pip\psp)=\bz$. The statistical
significance of this signal is $6.4\,\sigma$; the significance
including systematic uncertainty from the fit models is $5.4\,\sigma$.
These results agree with, and supersede previous measurements based on
the same data sample reported in Ref.~\cite{Z4430}.

In addition we determine the branching fraction
$\br(B^0\rt\kst(892)^0\psp)=\bkstpsp$ and the fraction of $\kst(892)$
mesons that are longitudinally polarized $f_L=\fl$.  These are the
first measurements of these quantities that are derived from a Dalitz
plot analysis.

\begin{center}
{\bf Acknowledgments}
\end{center}
We thank the KEKB group for excellent operation of the accelerator,
the KEK cryogenics group for efficient solenoid operations, and the
KEK computer group and the NII for valuable computing and SINET3
network support.  We acknowledge support from MEXT, JSPS and Nagoya's
TLPRC (Japan); ARC and DIISR (Australia); NSFC (China); DST (India);
MEST, KOSEF, KRF (Korea); MNiSW (Poland); MES and RFAAE (Russia); ARRS
(Slovenia); SNSF (Switzerland); NSC and MOE (Taiwan); and DOE (USA).


\begin{thebibliography}{99}
\bibitem{conjugate} The inclusion of charge-conjugate modes is implied
  throughout this paper. By $\psp$ we mean the $\psi(3686)$, also
  known as the $\psi(2S)$.
\end{thebibliography}
\end{document}